\documentclass[longauth]{aa}

\usepackage{graphicx}
\usepackage{color}
\usepackage{txfonts}
\usepackage{hyperref}
\usepackage{soul}
\usepackage{url}

\usepackage{natbib}
\bibpunct{(}{)}{;}{a}{}{,}

\usepackage[switch]{lineno}
\linenumbers

\usepackage{xcolor}
 
\begin{document}

   \title{MAGIC's interferometric diameters of \texorpdfstring{$\theta$}{0} Ophiuchi and \texorpdfstring{$\beta$}{b} CMa and comparison with asteroseismic modelling}
   \titlerunning{Angular diameters of \texorpdfstring{$\beta$}{b} Cephei stars and asteroseismic modelling}

\author{
K.~Abe\inst{1} \and
J.~Abhir\inst{2} \and
A.~Abhishek\inst{3} \and
V.~A.~Acciari\inst{4} \and
A.~Aguasca-Cabot\inst{5} \and
I.~Agudo\inst{6} \and
I.~Albanese\inst{7} \and
T.~Aniello\inst{8} \and
C.~Arauner\inst{9} \and
A.~Arbet Engels\inst{10}
% ...
}
\institute { Japanese MAGIC Group: Department of Physics, Tokai University, Hiratsuka, 259-1292 Kanagawa, Japan
\and ETH Z\"urich, CH-8093 Z\"urich, Switzerland
\and Universit\`a di Siena and INFN Pisa, I-53100 Siena, Italy
\and Institut de F\'isica d'Altes Energies (IFAE), The Barcelona Institute of Science and Technology (BIST), E-08193 Bellaterra (Barcelona), Spain
\and Universitat de Barcelona, ICCUB, IEEC-UB, E-08028 Barcelona, Spain
\and Instituto de Astrof\'isica de Andaluc\'ia-CSIC, Glorieta de la Astronom\'ia s/n, 18008, Granada, Spain
\and Universit\`a di Padova and INFN, I-35131 Padova, Italy
\and National Institute for Astrophysics (INAF), I-00136 Rome, Italy
\and Technische Universit\"at Dortmund, D-44221 Dortmund, Germany
\and Max-Planck-Institut f\"ur Physik, D-85748 Garching, Germany
% ...
}

\date{Received 15/06/2026; accepted 27/10/2026}

  \abstract
    {$\beta$ Cephei type stars are massive ($8-25$M$_{\odot}$) and known to have several pulsation modes with periods typically of a few hours. Despite a better understanding of their physics since 1992, the inner structure of these stars involves key stellar physics processes that are still uncertain. Asteroseismology offers insights into these variable stars by analysing pulsation modes, helping to probe their inner layers and understand their structure and mixing processes.}
    {Building an asteroseismic model that would reliably predict an observable of a $\beta$ Cephei star would be a significant improvement, thereby reducing the uncertainty on their internal structure (e.g., overshooting parameter $\alpha$ or metallicity).}
    {We measure the angular diameter of two $\beta$ Cephei stars, $\theta$ Oph and $\beta$ CMa, using the MAGIC Stellar Intensity Interferometer and compare them, together with the Gaia and Hipparcos parallaxes, to the predictions of the asteroseismic model. The latter are based on Levenberg-Marquardt local minimisation of stellar evolution model parameters to fit observed pulsation modes.}
    {In the case of $\theta$ Oph, the observed angular diameters are compatible with the asteroseismic model. For $\beta$ CMa, the radius derived from the angular diameter and the Hipparcos parallax is in tension, at the $2.5\sigma$ level, with the asteroseismic prediction. We show that this tension mirrors a pre-existing conflict between independent spectroscopic determinations of the surface gravity of the star, that percent-level systematics of the stellar models are quantitatively insufficient to explain it, and that it most likely points to a bias of the Hipparcos parallax. In the frame of this paper, the statistical uncertainty on the interferometric results is similar to the systematic uncertainty on the parallaxes.}
   {
   With the already implemented addition in early 2024 of the first Large-Sized Telescope of the northern hemisphere array of the Cherenkov Telescope Array Observatory (CTAO-North LST-1) to the setup, the uncertainty in the angular diameter of these stars is expected to decrease by an order of magnitude, allowing to further constrain stellar evolution models and parallaxes.}


\maketitle
\nolinenumbers

\section{Introduction}
    Asteroseismology is one of the few methods that can constrain the inner structure of stars. Analogously to how seismologists analyse sound waves from earthquakes, asteroseismology allows us to determine the density and composition of the stellar medium through which the oscillations propagate.
    
    Hence, this method is especially useful in the study of pulsating stars such as $\beta$ Cephei variables. These stars are massive, and their brightness can change by 0.01 to 0.3 magnitudes over timescales of a few hours. 
    \cite{cox_opacity_1992} explained the pulsation phenomenon with the $\kappa$ mechanism, which is the preferred scenario to date.
    Usually, the opacity $\kappa$ of a star decreases as it contracts and heats, following Kramers' law $\kappa \propto \rho/T^{3.5}$. However, in $\beta$ Cephei, inner layers have a high iron density and at a temperature of $\sim 200\,000$~K, partial ionisation zones can form. As a result, less energy passes through the layer, increasing the opacity. Hence, the pressure rises, and the layer overshoots its equilibrium position. It follows that the pressure and temperature decrease, which also decreases the opacity, and the star starts a new cycle \citep{goss_asteroseismology_2012}. 
    These processes are challenging to model and their understanding could refine stellar evolution models as they depend on quantities such as the overshooting parameter \citep{pedersen_shape_2018}, which impacts the mixing of layers, providing more fuel to the stellar core \citep{claret_dependence_2016}.

    A way to determine the accuracy of an asteroseismic model is to give it observational parameters, such as the chemical composition or the pulsation modes, and derive a quantity that can be measured, such as the stellar radius.

    The MAGIC telescopes are two Cherenkov Telescopes located in the Roque de los Muchachos Observatory, La Palma, Spain. While they were originally built to perform very-high-energy gamma-ray observations, they were slightly modified to also perform as a stellar intensity interferometer (MAGIC-SII) \citep{magic_2024}. In this mode of observation, the setup is used to measure stellar angular diameters of the order of 1 milliarcsecond in B-band with a $\sim10$~\% precision in about 2.5 h for stars up to magnitude 4. Thus, these measurements, together with distance estimates from other experiments (Gaia/Hipparcos), can be compared with the stellar radii predicted by asteroseismic models, hence testing the models themselves.

    In this paper, we compare simulations and MAGIC-SII results for two $\beta$ Cephei variables: $\theta$ Oph and $\beta$ CMa (Mirzam).
    $\theta$ Oph is a $9$~M$_{\odot}$ star with a V-magnitude varying between 3.26 and 3.29 \cite{van_hoof_brightness_1962}, making it a suitable target of study for MAGIC-SII. This star is visible for a significant part of the year, allowing frequent observations \citep{wenger_simbad_2000}. 
    Mirzam has similar characteristics as it is a $\sim12$~M$_{\odot}$ star with a magnitude that varies between $1.97$ and $2.01$ over $6$ hours \citep{wenger_simbad_2000}. 
    Neither star is classified as a fast rotator, as their projected rotational velocities ($v\sin\,i$) are around $30$~km/s \citep{briquet_asteroseismic_2005, mazumdar_asteroseismic_2006}.

    We use MAGIC-SII capabilities to derive the stellar radius for $\theta$ Oph and $\beta$ CMa, incorporating Hipparcos and Gaia parallaxes when available. These results are then compared with predictions from an asteroseismic model.
    
    This paper is structured as follows. Section \ref{sect:SII} introduces the intensity interferometry technique and how it is used to measure the angular diameter of a star. Sections \ref{sect:MAGIC_setup}  and \ref{sect:MAGIC_results} respectively describe the MAGIC-SII instrumental setup and data analysis. The asteroseismic model and its parameters are detailed in section \ref{sect:seismic_model}. Comparisons between the two methods along with previous results are discussed in section \ref{sect:discussion}. Section \ref{sect:conclusion} summarises the main conclusions of this work.

\section{SII measurements} \label{sect:SII}
    The measurement of stellar angular diameters using SII relies on two physical effects: temporal and spatial correlations, both of which are described in this section.
    
    \subsection{Temporal correlation function}
    The temporal correlation function provides a statistical description of the observed light. Its first application in astronomy was by \cite{hanbury_brown_test_1956}, who developed and used this technique to measure the angular diameter of Sirius. Given two detectors (labelled 1 and 2), the second-order correlation function $g^2(\tau)$ is defined as follows:

    \begin{equation}
        g^2(\tau) = \dfrac{\langle I_1(t) \cdot I_2(t+\tau) \rangle}{\langle I_1(t) \rangle \, \langle I_2(t) \rangle} \; , 
    \label{eq:g2}
    \end{equation}
    
    \noindent where $I_i(t) = |E_i(t)|^2 $ is the intensity as a function of time, and $E_i(t)$ is the complex electric field measured by the $i^{\text{th}}$ detector. $\langle.\rangle$ denotes the time average, satisfying the property $\langle I_i(t) \rangle=\langle |E_i(t)|^2 \rangle=\langle |E_i(t+\tau)|^2 \rangle$. The second-order correlation function is used to probe the quantum properties of light as it entirely depends on the source. For thermal light (characterised by random phase) $g^2(0)=2$,  in which case the Siegert’s relation holds: $g^2(\tau)=1 + |g^{1}(\tau)|^2$, where $g^1(\tau)$ is the first-order correlation function. In practice, $g^2(\tau)$ is obtained by correlating the arrival times of photons detected by two detectors. This phenomenon is also known as "photon bunching".
    
    \subsection{Spatial correlation}
    To measure spatial correlation, the same source must be observed from two different locations, and the arrival times of photons at these positions must be correlated. The temporal correlation measured in this process is modulated by the spatial correlation function, which encodes key information about the star's physical properties. Specifically, the spatial correlation function is directly related to the source’s brightness distribution, and the observed correlation is proportional to the Fourier transform of that distribution.
    For a uniform disk (UD), the spatial correlation function is given by:

    \begin{equation}
        |V_{12}(d)| = 2\dfrac{J_1(\pi\,d\,\theta_\text{UD}/\lambda)}{\pi\,d\,\theta_\text{UD}/\lambda}.
    \label{eq:V2}
    \end{equation}

    $J_1$ is the first-order Bessel function, $\theta_\text{UD}$ is the angular diameter of the star, $d$ is the projected baseline between two detectors (telescopes) and $\lambda$ is the observed wavelength. Since the temporal correlation is modulated by the spatial one, the overall measured correlation follows $g^2(\tau, d) = 1 + |g^1(\tau)|^2 \cdot |V_{12}(d)|^2$, which implies that measuring the angular diameter requires correlating the time of arrival of photons across different baselines.

\section{MAGIC-SII measurements and analysis}\label{sect:MAGIC_setup}
    
    The MAGIC Imaging Atmospheric Cherenkov Telescopes (IACTs) are located in La Palma, Spain. 
    With 17-metre diameter reflector dishes, MAGIC 1 and MAGIC 2 (M1 and M2) are separated by 85 metres and were commissioned in 2004 and 2009, respectively. The first Stellar Intensity Interferometry (SII) observations were conducted in 2019 \citep{acciari_optical_2020}.

    Since very-high-energy gamma-ray observations are MAGIC's primary objective, MAGIC-SII measurements are mainly performed during the bright moon period. In 2022, 192 hours of data were recorded and analysed \citep{magic_2024}. As a result, 22 stellar diameters were published, nine of which were compared with previous results obtained by other SII observatories, demonstrating good accuracy.

    As MAGIC's SII capabilities were already extensively described \citep{acciari_optical_2020,magic_2024}, this section focuses only on relevant details for this work.
    
\subsection{Optics and detection}
    
    The MAGIC camera consists of 1039 pixels, and each one consists of a Winston cone that collimates the light into fast photomultiplier tubes (PMTs). Because the telescopes are designed to observe Cherenkov light, the PMTs were specifically chosen for their high quantum efficiency in the blue light regime.

    MAGIC-SII only uses 3 pixels of the camera. In a few seconds, a filter holder with 6 slots can be deployed in front of the camera, allowing SII measurements. Currently, three band-pass filters (centred at $425$~nm with a FWHM of $26$~nm) are mounted on the holder, and the remaining slots are closed with an opaque plastic cap.
    The Active Mirror Control (AMC) allows light to be focused into any of the six slots in the filter holder. 
    In the "Full Mirror" configuration, all the incoming light can be directed onto a single pixel. The PMTs are connected in parallel to two different measurement setups:

    \begin{itemize}
        \item The slow data acquisition provides low-frequency ($1$~Hz) monitoring of the PMT anode current and high voltage. It is primarily used to detect slow rate variations in the anode current due to clouds, wind, or any other slow variation.

        \item The fast data acquisition transmits the signal over 162 metres of optical fibre to the Femto HSA-Y-2-40 amplifier, which is connected in series to two Spectrum M4i-4450-x8 digitisers. This setup enables a data acquisition rate of $500$~MSamples/s, corresponding to a time resolution of 2 ns. The data is then processed by a GPU, which computes correlations in real-time.
    \end{itemize}  
    
    Since the photon detection devices used are PMTs, the exact number of detected photons (\textit{ie.} the intensity $I_i$) is not directly measurable and must be reconstructed using the PMT detection properties. 
    The slow measurement of the anode current ($DC_i$) provides the average intensity $\langle I_i \rangle$, which includes contributions from the star signal as well as other background sources (primarily night sky background (NSB), moonlight, and potential stray light).
    On the other hand, the fast measurements are the variations around the average. Since the statistics follow a Poisson distribution, the fluctuations are proportional to the square root of the $DC$ measurements $\sqrt{\langle (I_i(t)-\langle I_i(t)\rangle)^2\rangle}\propto \sqrt{DC_i} $, where the index $i$ labels the telescope ($i=1$ for M1).

\subsection{Correlation}\label{subsect:correlation}

    A convenient quantity derived from $g^2(\tau, d)$ is the normalised contrast $c(\tau, d)= |g^1(\tau)|^2 \cdot |V_{12}(d)|^2$ \citep[eq.4, 6]{acciari_optical_2020}.
    By dividing the normalised contrast at a given baseline by its value at zero baseline, one recovers the squared visibility $|V_{12}(d)|^2$. The contrast $c(\tau=0, d=0)$ only depends on the telescope system and not on the observed target and is often referred to as the zero baseline correlation (ZBC). 
    
    To perform a ZBC calibration, a bright, single star with an angular diameter $\sim1$mas must be analysed. Adhara ($\epsilon$ CMa) is a suitable star and was used as the ZBC calibrator for the MAGIC-SII analysis \citep{magic_2024}.

    Although \cite{magic_2024} does not find significant evidence of the time evolution of the ZBC value, more recent analyses covering a larger time range show a slow evolution, potentially due to PMT degradation. Hence, for both the ZBC and the stellar measurements, we divided the data into different observing periods and performed independent analyses on each of them (see figure \ref{fig:MAGIC_obs_time}). While this approach reduces the amount of data taken into account in each analysis, it ensures that the result remains independent of ZBC evolution.
    
    Numerically, the correlation is computed using Pearson's correlation coefficient $\rho(\tau)$ and the normalised contrast is obtained: 
    
    \begin{equation}
        c(\tau, d) \propto \dfrac{ \rho(\tau)} {\sqrt{DC_1 \cdot DC_2}}
    \label{eq:correlation}
    \end{equation}
    
    The full relation \citep{magic_2024} includes corrections for PMT high-voltage variations and the ratio between the stellar and NSB fluxes.
    A $12$~MHz high-pass filter is applied to $c(\tau, d)$ to remove slow correlated electronic noise. The maximum of the correlation peak is found by fitting a Gaussian function filtered with the same high-pass filter. 
    
    The expected delay for the correlation peak $\tau_0$ is obtained from the projected baseline derived from the relative positions of the two telescopes and the direction of the target. 
    This allows for the derivation of an averaged normalised contrast for each projected baseline, which can be plotted as a function of the projected baseline as in \cite[figure 6]{magic_2024}.
    Once the average normalised contrast is obtained for each projected baseline and the ZBC has been obtained, equation \ref{eq:V2} allows to reconstruct the angular size of the star $\theta_{UD}$.

\section{Analysis on \texorpdfstring{$\theta$}{0} Oph \& \texorpdfstring{$\beta$}{b} CMa}\label{sect:MAGIC_results}
    
    \begin{figure}
      \resizebox{\hsize}{!}{\includegraphics{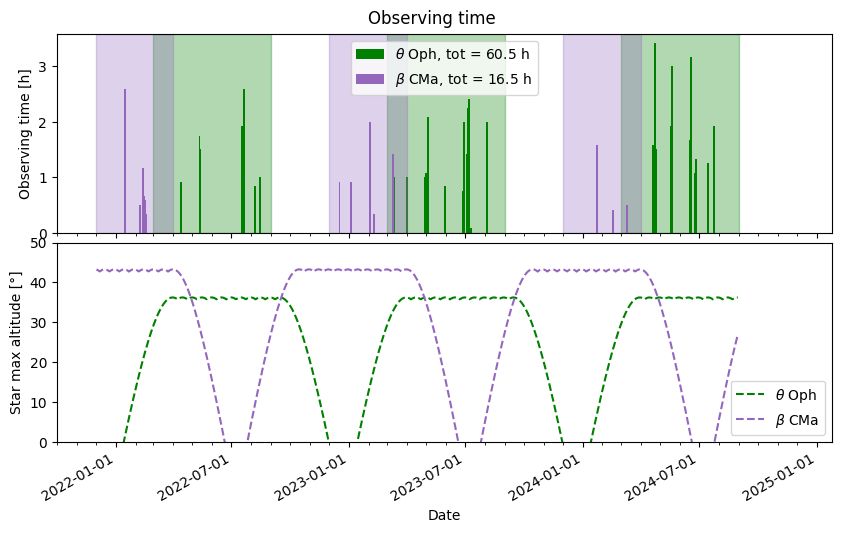}}
      \caption{Time of observation for $\theta$ Oph and $\beta$ CMa from 2021 to the end of August 2024. $\theta$ Oph was more observed than $\beta$ CMa, particularly in the last two years. The vertical span represents the time ranges on which the analyses were performed. The bottom graph shows the maximum altitude per night.}
      \label{fig:MAGIC_obs_time}
    \end{figure}

    Observations of the two stars began in 2021 (end of January for $\beta$ CMa and end of June for $\theta$ Oph). The analysis uses data extending until the end of 2024 (see figure \ref{fig:MAGIC_obs_time}).
    
    For data selection, the analysis software automatically makes a selection based on filenames, discarding specific files such as those taken for specific tests. Additionally, users can apply custom quality cuts. In this analysis, the following cuts were applied:

    \begin{itemize}

        \item  The first filter allows the inclusion/exclusion of given time periods. For the $\theta$ Oph analysis, we used available data from January 2022 to August 2024 with yearly time cuts ranging from March to September. For Mirzam, we used data from January 2022 to October 2024, considering only data from December to April.
        This filter can, for instance, be used to verify that the ZBC does not evolve by performing a time-dependent analysis of a calibrator. 
        
        \item The second filter ensures that the HBT S/N for a given baseline is greater than 3$\sigma$ to guarantee that the signal is strong enough for a reliable analysis.
        
        \item Additionally, when the peak is fitted with a Gaussian (using the $12$~MHz high-pass filter), we verify that the corrected delay is below $1.5$~ns, since the closer to 0~ns the better is the calculation of the expected delay for the correlation peak $\tau_0$ (see section \ref{subsect:correlation}).
        
        \item Finally, a cut on the peak width is applied. The Gaussian standard deviation should not be larger than $2.95$~ns. As the expected correlation shape has a $\sigma\sim2.2$~ns \citep[sect.3.2]{magic_2024}, the cut ensures that the timing accuracy is constrained but also verifies that the delay for all sub-measurements is properly corrected.
       
    \end{itemize}

\begin{figure}
      \resizebox{\hsize}{!}{\includegraphics{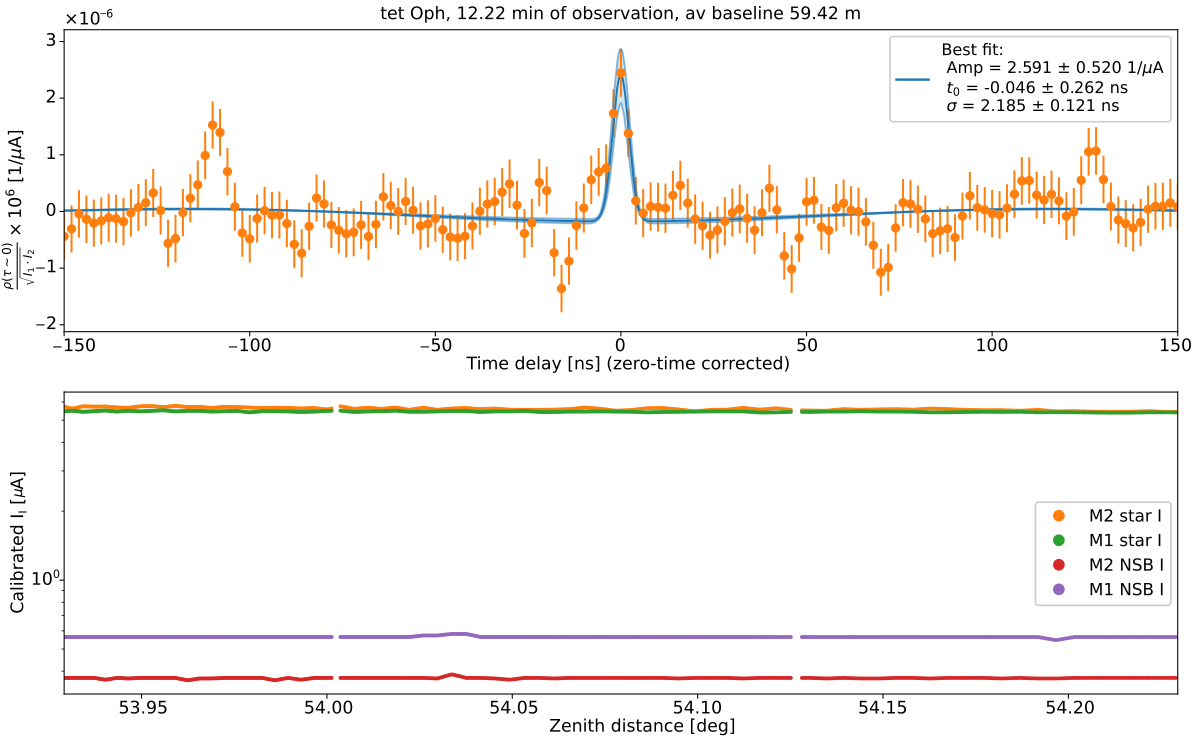}}
    \caption{
      (\textit{top}): Normalised contrast of $\theta$ Oph (orange) for a baseline bin of $5$~m centred on $59.10$~m over $12.22$~min of observation. Also shown in blue is the Gaussian fit using a $12$~MHz high-pass filter.
      (\textit{bottom}): DC currents for the star signal and NSB, for the two telescopes M1 and M2. Changes are visible in both the star signal and NSB, which are taken into account.}
      \label{fig:MAGIC_TetOph_72.4}
    \end{figure}

\begin{figure*}
    \centering
       \includegraphics[width=17cm]{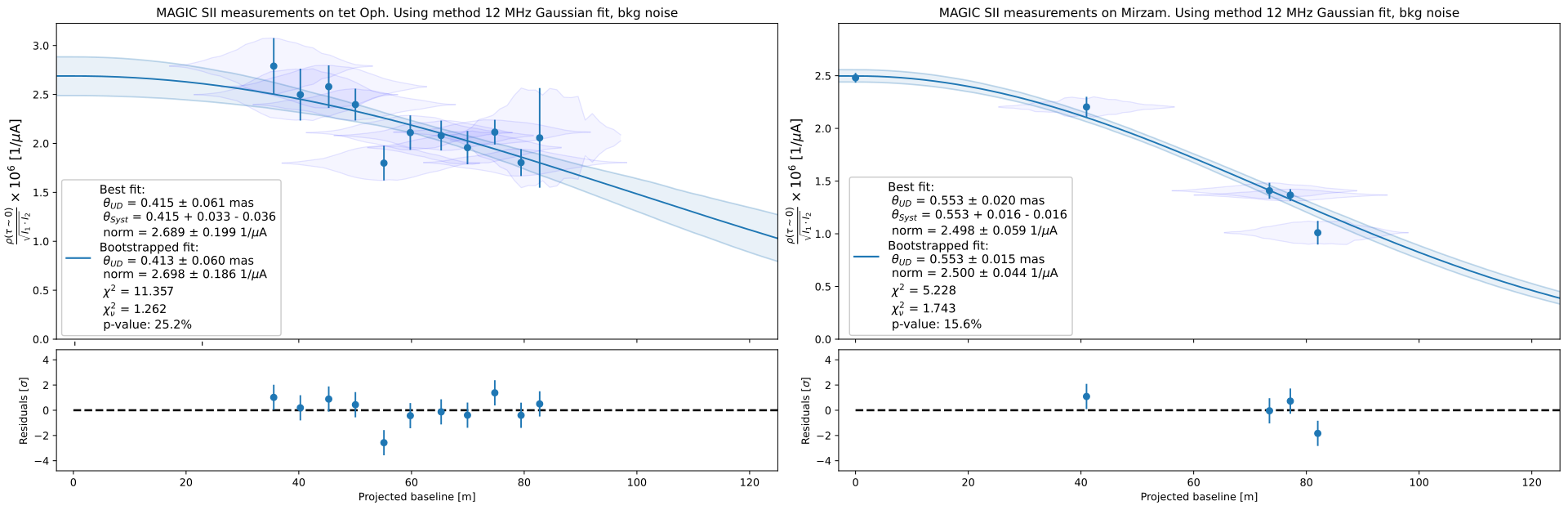}
         \caption{\textit{left:} Squared visibility curve for $\theta$ Oph obtained for the 2024 period. No ZBC correlation point was added.
         \textit{right:} Squared visibility curve for Mirzam for the period from December 2023 to April 2024. The ZBC correlation point is derived from an analysis of Adhara, selecting data from the same period to account for ZBC evolution. Its value is $2.420 \pm 0.071$~/$\mu$A.}
         \label{fig:MAGIC_TetOph_VisCurve}
    \end{figure*}

    For this analysis, we used fixed 5~m baseline bins, following the approach explained in the MAGIC-SII performance paper \citep{magic_2024}. After applying the cuts and filters, the observed time per baseline bin ranges from 12 to 256 minutes for $\theta$ Oph and from 10 to 134 minutes for Mirzam. For $\theta$ Oph, even in the bin with the smallest exposure (roughly $12$ minutes), the signal-to-noise ratio is sufficient for a good fit using a Gaussian with a $12$~MHz high-pass filter (see figure \ref{fig:MAGIC_TetOph_72.4}). It is also the case for $\beta$ CMa. The noisy baseline can be an issue for fainter stars. 
    The squared visibility curve for the 2024 observing periods is shown in figure \ref{fig:MAGIC_TetOph_VisCurve}. 

\subsection{Assumptions} 

\begin{figure}
      \resizebox{\hsize}{!}{\includegraphics{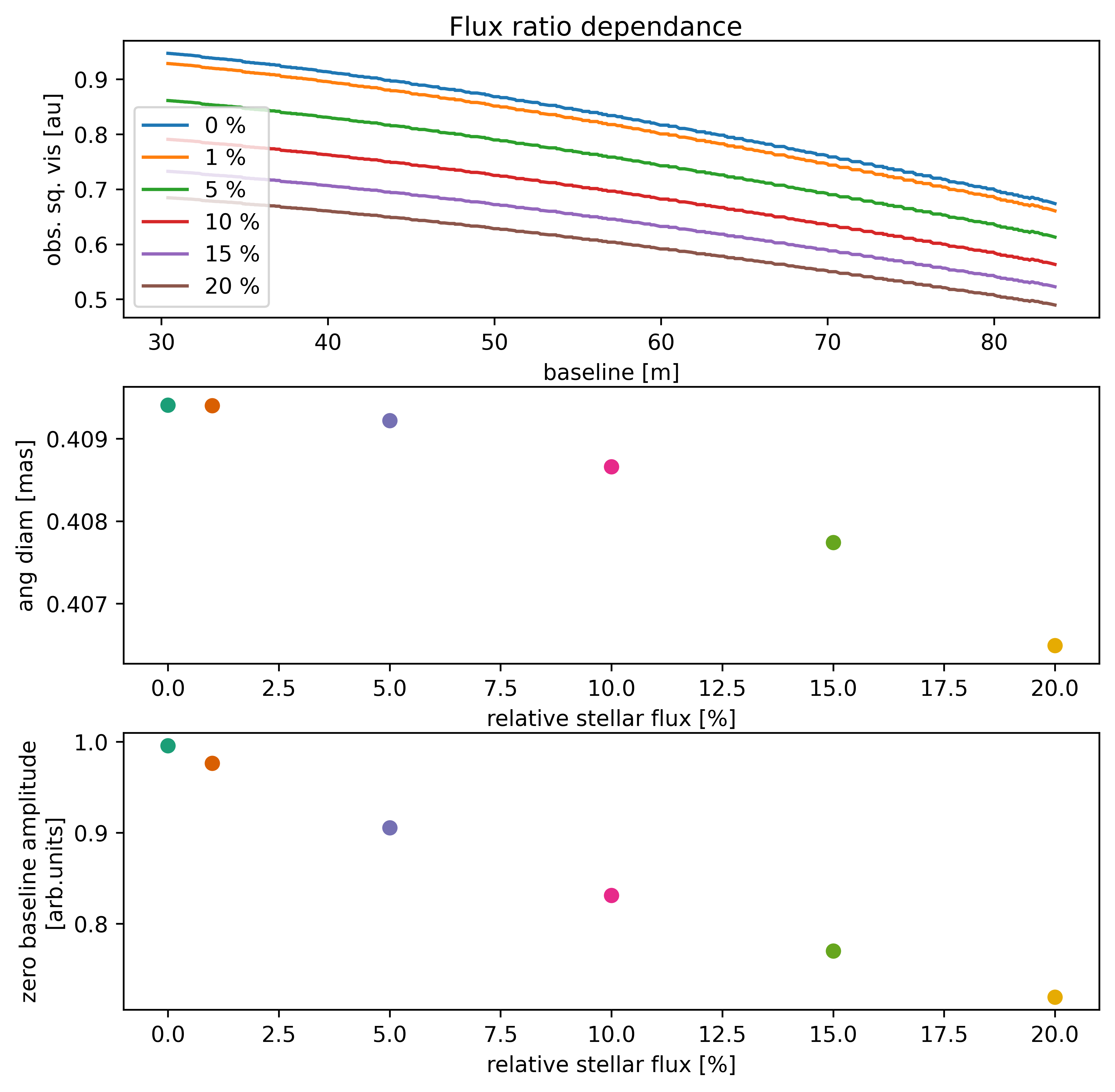}}
      \caption{Analysis of the effect of the $\theta$ Oph B (for different relative stellar fluxes) on the measured squared visibility as a function of baseline (\textit{top}), the impact on the derived angular diameter (\textit{centre}) and on the amplitude of the squared visibility at zero baseline (\textit{bottom}), as a function of the relative stellar fluxes. 
      }
      \label{fig:re_flx_study}
    \end{figure}

    $\theta$ Oph is a triple system. The primary component ($\theta$ Oph Aa) is the $\beta$ Cephei ($V_{mag}=3.248$, B2IV type) that we are interested in \citep{briquet_asteroseismic_2005}. The close component discovered in 2005 ($\theta$ Oph Ab) is a $<1$~M$_\odot$ SB1 star with an orbital period of 56.71 days and an eccentricity of 0.1670 \citep{briquet_asteroseismic_2005}. 
    The third star ($\theta$ Oph B) is a main-sequence B5 speckle companion with $V_{mag} = 4.87$ \citep{walczak_seismic_2019}. It has an orbital period of approximately 100 years and is separated from the primary by $0.15''$ \citep{briquet_asteroseismic_2005, shatsky_mass_2002}. More recent studies suggest that the period is significantly shorter than a century, and around 14 years \citep{walczak_seismic_2019}. 
    Our first assumption was to ignore the close $\theta$ Oph Ab component due to its higher magnitude and redder spectrum, which has a negligible effect in our wavelength range. 
    $\theta$ Oph B, which has a magnitude closer to that of $\theta$ Oph Aa, is also a B-type star, which could impact measurements and analyses. Therefore, we studied the effect of this star on the expected squared visibility curve for various parameters such as the flux ratio (the flux of $\theta$ Oph B over the total flux of the system), angular separation, and moment of observation. 
    We also accounted that its radius is $80\%$ of that of the primary B2 star \citep{pecaut_intrinsic_2013}. 
    We estimate the error on the predicted angular diameter by fitting the simulated squared visibility with a uniform disk model. The simulated squared visibility was obtained from \cite{berger_introduction_2007} and is convolved with the MAGIC mirror shape to reproduce the measurements. We find that for flux ratios between 5\% and 20\% \citep{walczak_seismic_2019, shatsky_mass_2002}, the error on the angular diameter is $\leq 1$\%. The error is comparable if the angular separation between the two companions is greater than $20$~mas. The error on the zero baseline amplitude and angular diameter varying the relative flux is shown in figure \ref{fig:re_flx_study}.
    While the error on the predicted angular diameter is below 1\%, the error on the amplitude of the zero baseline correlation can vary up to $\sim20$\% in some cases. Therefore, without any precise knowledge of the system, we cannot use the zero baseline correlation from Adhara or any other calibrator. Consequently, we anticipate a rather large angular diameter uncertainty.

\subsection{Results}\label{sect:results}

\begin{figure}
      \resizebox{\hsize}{!}{\includegraphics{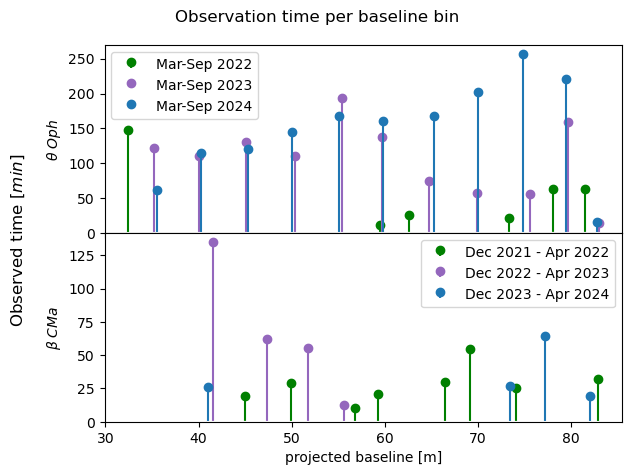}}
      \caption{Time of observation for $\theta$ Oph and Mirzam per baseline bin. The integrated signal for the baseline bin with the least exposure was good enough to perform a good fit.}
      \label{fig:MAGIC_baseline_time}
    \end{figure}

    We processed the data using the \texttt{magic-spysii} software developed by the MAGIC-SII team and applied the quality checks described above. The integrated time per baseline is shown in figure \ref{fig:MAGIC_baseline_time}. 
    For each star, three sub-analyses are performed, and the overall angular diameter is determined by performing a combined analysis. This takes into account that the individual results represent the same star and hence assumes that they have the same angular diameter. It provides smaller uncertainties than would be obtained by performing a direct average.
    
    Table \ref{table:results} presents the resulting angular diameters, along with statistical and systematic uncertainties, both calculated using the analysis software. Note that the $\beta$ CMa result slightly differs from the one obtained in \cite{magic_2024}. This difference can be explained by the use of the chessboard mirror configuration, which represents about $50$~\% of the 2022 observations and was not taken into account in this work.

    To derive a stellar radius from the measured angular diameter, we use the parallax $\bar{\omega}$ from Hipparcos and Gaia when available. However, this approach often results in significant uncertainties, as the uncertainty on the distance is proportional to $1/\bar{\omega}^2$. 
    The Hipparcos re-reduction \citep{vanleeuwen07} reported a parallax of $7.48 \pm 0.17$~mas for $\theta$ Oph and $6.62\pm0.22$~mas for $\beta$ CMa. We note that the original Hipparcos catalogue \citep{perryman97} gives $6.53\pm0.66$~mas for $\beta$ CMa: the two reductions agree in value, but their quoted uncertainties differ by a factor of three, a point we return to in Sect.~\ref{sect:discussion}.
    Gaia reported a value of $9.79\pm1.18$~mas for $\theta$ Oph parallax, which comes from the second data release (DR2), published in 2018 \footnote{See \url{https://www.cosmos.esa.int/web/gaia/dr2}}.  
    Gaia DR2 parallaxes for bright targets have signs of large bias ($\sim0.5-1$~mas for $G\leq3$) \cite{drimmel_possible_2019}, which seems to be also the case for Hipparcos parallaxes \cite{pan_systematic_2007}.
    
    The error propagation accounts for both the parallax uncertainty and the statistical and systematic uncertainties on the angular diameter. The relative uncertainty on the inferred radius is given by: 
    
    \begin{equation}
        \dfrac{\delta R_{\star}}{R_{\star}} = \sqrt{
        \left( \dfrac{\delta \bar{\omega}}{\bar{\omega}} \right)^2
        +
        \left( \dfrac{\delta \theta_\text{UD}}{\theta_\text{UD}} \right)^2}\; . 
    \end{equation}

    \noindent  The $\delta$ represents the uncertainty of the value. The angular diameter uncertainty is an output of the \texttt{magic-spysii} software.
    
    Since parallax uncertainty often leads to rather large uncertainties, obtaining a precise parallax measurement is crucial. If the parallax uncertainty is too large, additional MAGIC-SII data collection does not improve the final accuracy, as the total uncertainty remains dominated by the parallax contribution. However, this is not the case since the uncertainty component due to the angular diameter is similar to the parallax one for the results of $\theta$ Oph assuming Gaia parallax. Their ratio increases to 5.6, assuming a Hipparcos parallax, indicating that further improvement is still possible on the MAGIC-SII uncertainty.

\begin{table}
    \caption{Uniform disk stellar angular diameter results.}
        \label{table:results}
        \centering
        \begin{tabular}{c c c}
        \hline
        Period & $\theta_{UD}$ [mas] & $\delta\theta_{UD}$ and $\delta_{sys}$ [mas]\\
        \hline\hline
        \multicolumn{3}{c}{$\theta$ Oph} \\
        \hline
        Mar 22 - Sept 22 & 0.474 & $0.129\pm0.018$ \\
        Mar 23 - Sept 23 & 0.334 & $0.114\pm0.040$ \\
        Mar 24 - Sept 24 & 0.415 & $0.061\pm0.035$ \\
        combined & \multicolumn{2}{c}{$0.400\pm0.051$} \\
        \hline\hline
        \multicolumn{3}{c}{$\beta$ CMa} \\
        \hline
        Dec 21 - Apr 22 & 0.509 & $0.049\pm0.018$ \\
        Dec 22 - Apr 23 & 0.599 & $0.097\pm0.030$ \\
        Dec 23 - Apr 24 & 0.553 & $0.020\pm0.016$ \\
        combined & \multicolumn{2}{c}{$0.547 \pm 0.016 $} \\
        \hline\hline
\end{tabular}
\end{table}

\section{Asteroseismic modelling}\label{sect:seismic_model}
$\beta$~CMa, together with $\beta$ Cephei, is a prototype of the $\beta$ Cephei pulsating class \citep[see a review of their properties in][]{stankov05}. The star has been observed for a long time, but it has not received a lot of attention from an asteroseismic point of view; its first main seismic modelling was carried out by \cite{mazumdar_asteroseismic_2006}. $\theta$ Oph was followed by a coordinated photometric and spectroscopic campaign from the ground by \cite{handler05} and \cite{briquet_asteroseismic_2005}, leading to its asteroseismic modelling in  \cite{briquet09}. The star is an important target among known $\beta$ Cephei pulsators because two rotational frequency splittings were detected, giving a constraint on its internal rotation profile. Furthermore, the stellar modelling revealed a surprisingly large value of the core overshooting, $\alpha_{\textrm{ov}}=0.44\pm0.07$, at odds with the fact that the star is a relatively slow rotator ($v \sin i \simeq 30$~km/s). In a study focused on non-adiabatic aspects of pulsations, \cite{jagoda09} also reported large values of overshooting, from 0.35 up to 0.55. In a less detailed modelling, though focused on the structural effects of rotation, \cite{lovekin10} found a lower value $\alpha_{\textrm{ov}}=0.28\pm0.05$. The star was recently observed by the BRITE space mission. With the new data, \cite{walczak_seismic_2019} observed the same $\beta$~Cep modes as from the ground, as well as 19 newly detected frequencies, most of them in the low-frequency regime of SPB pulsations. Most of the new seismic models derived in \cite{walczak_seismic_2019} also present $\alpha_{\textrm{ov}}$ between 0.28 and 0.40. With the new information delivered by the interferometric radius, an ideal 
time has come to reinvestigate the two stars.\\

We have thus adapted the local Levenberg-Marquardt algorithm that we used for modelling solar-like stars \citep{miglio05,salmon21} to the specific case of $\beta$~Cep stars and carried out a new asteroseismic modelling of the two stars. \cite{salmon22} showed that the stellar mean density is inferred from $\beta$~Cep pulsations with almost no model dependence. Consequently, we compare the radii predicted by the asteroseismic forward approach with those derived from the interferometric angular diameters and the available parallaxes. In addition, since the stellar mean density is the quantity most directly and most robustly constrained by the pulsations, we also combine it with the interferometric radius to derive an independent mass estimate, providing a quasi-model-independent consistency test (Sect.~\ref{sect:discussion}).\\ 

\subsection{Observational dataset: \texorpdfstring{$\beta$}{b}~CMa}
The most complete asteroseismic dataset for $\beta$~CMa was presented in \cite{mazumdar_asteroseismic_2006}. The authors analysed 4.5 years of spectroscopic observations and detected three pulsation frequencies for which mode identification was obtained, though one of them is less firm. Their seismic dataset and other observational constraints are given in Table~\ref{table-prop-bcma}. The mode with frequency $\nu_1$ is a radial pulsation, which the authors identified most likely as the first overtone since only theoretical models with radial order $n=2$ for this mode can reproduce the other observational constraints. The asteroseismic modelling also suggested $\nu_3$ is the gravity mode $n=-1$, with angular degree $\ell=3$, although this complete identification remains subject to caution \citep[see][]{mazumdar_asteroseismic_2006}. In our analysis of the star, we did two tests: at first, using a constraint set with the three reported identifications, and a second one where we discarded the $\nu_3$ mode.\\

 \begin{table*}
 \caption{Observed frequencies ($\nu$), effective temperature (T$_{\textrm{eff}}$), surface gravity ($\log g$), metallicity (Z), equatorial velocity ($v_{\textrm{eq}}$) for $\beta$~CMa.}
 \label{table-prop-bcma}
 \begin{center}
 \begin{tabular}{lll}
 \hline
 Observable & Value & Reference \& Comment \\
 \hline \hline
 $\nu_1(\ell=0,n=1;2)$ [$\mu$Hz] & 46.291 $\pm$ 0.001 & \cite{mazumdar_asteroseismic_2006}; id. by LPV; likely $n=2$  \\ 
 $\nu_2(\ell=2,m=2,n=-1)$ [$\mu$Hz] & 46.057 $\pm$ 0.001 & \cite{mazumdar_asteroseismic_2006}; id. by LPV   \\ 
 $\nu_3(\ell=3,m=2,n=-1)$ [$\mu$Hz] & 48.417 $\pm$ 0.001 & \cite{mazumdar_asteroseismic_2006}; id. a posteriori from modelling  \\ 
 T$_{\textrm{eff}}$ [K] & 24000 $\pm$ 1000 & \cite{morel06}\\
 $\log g$ [cm/s$^2$]& 3.50 $\pm$ 0.15 & \cite{morel06} \\
 $Z$ & 0.0104 $\pm$ 0.0021 & \cite{morel06} \\
 $v_{\textrm{eq}}$ [km/s] & 31 $\pm$ 5& \cite{mazumdar_asteroseismic_2006}; moment method on LPV  \\
 
  \hline
 \hline
 \end{tabular}
 \end{center}

 \end{table*}

Since the $\nu_2$ and $\nu_3$ modes are non-axisymmetric modes (their azimuthal order $m\neq0$), we corrected their frequencies by correcting for a mean rotation effect. We used for the corrective term a value of 1~$\mu$Hz, representative of the values found by \cite{mazumdar_asteroseismic_2006} for $m\beta v_{\textrm{eq}}/R$, where $\beta$ is the Ledoux constant \citep{ledoux51} of a given mode, and $R$ the radius of the models (computed during the asteroseismic modelling). This frequency shift for non-axisymmetric modes is obtained from a first-order perturbative treatment of the rotation effect on the pulsations. At this order, the physical effect induced by rotation is a contribution purely of the Coriolis force \citep[e.g.][]{goupil11}. Applying this correction gives the expected values of the $m=0$ modes, which are more directly obtained from the numerical computations of the pulsations of stellar models.\\

\subsection{Observational dataset: \texorpdfstring{$\theta$}{0} Oph}

Ground-based spectroscopic and multicolour photometric observation campaigns were dedicated to $\theta$ Oph, whose results were reported in \cite{handler05} and \cite{briquet_asteroseismic_2005}. The combined data resulted in the identification of seven pulsation modes, which are given in Table~\ref{table-prop-thetaoph}, alongside other constraints. The radial $\nu_1$ mode is identified most likely as the fundamental one, from a posteriori asteroseismic modelling of the star. The detection of two rotational multiplets makes it an important target to characterise the internal rotation of the star. The detection of prograde and retrograde modes in the $\ell=2$ triplet ($\nu_5,\nu_6,\nu_7$) allows to easily interpolate their frequencies to estimate the frequency of the $m=0$ mode. The observational errors on the frequencies are at most 0.006$\mu$Hz \cite{handler05}.\\

The system is triple, with a B5 spectral-type star appearing as a speckle binary of the primary spectroscopic binary. The star contributes by ~20\% to the intensity of the spectroscopic lines of the system. Two analyses of the spectra resulted in a difference of 1000 K in the determinations of the effective temperature T$_{\textrm{eff}}$ \citet{briquet_asteroseismic_2005, briquet09}.
A third independent T$_{\textrm{eff}}$ was derived in \citet{handler05}  based on narrow-band photometry. This latter value is lower by $\sim$1000~K and in agreement with a low-resolution spectral determination by \citet{niem05}. We report in Table~\ref{table-prop-thetaoph} the three values. We also give the surface gravities ($\log g$) from the studies which accounted for the contamination of the speckle binary companion. Although they are compatible by pairs within their error bars, the asteroseismic results favour the cooler T$_{\textrm{eff}}$ value found by \citet{handler05}, which we will adopt as our reference. \\

 \begin{table*}
 \caption{Same as in Table~\ref{table-prop-bcma} for $\theta$~Oph, except for the surface rotation given as the projected 
equatorial velocity ($v \sin i$).}
 \label{table-prop-thetaoph}

 \begin{tabular}{lll}
 \hline
 Observable & Value & Reference \& Comment \\
 \hline \hline
 $\nu_1(\ell=0,n=1;2)$ [c/d] & 86.432  & \cite{handler05}; id. by multicolour photometry;   \\ 
 & & likely $n=1$ \\
 $\nu_2(\ell=1,m=-1,n=1)$ [$\mu$Hz] & 89.883 & \cite{handler05}; id. by multicolour photometry   \\ 
 $\nu_3(\ell=1,m=0,n=1)$ [$\mu$Hz] & 91.137 & \cite{handler05}; id. by multicolour photometry  \\ 
 $\nu_4(\ell=1,m=1,n=1)$ [$\mu$Hz] & 92.285 & \cite{handler05}; id. by multicolour photometry \\ 
 $\nu_5(\ell=2,m=-1,n=-1)$ [$\mu$Hz] & 82.361 & \cite{briquet_asteroseismic_2005}; id. by LPV   \\ 
 $\nu_6(\ell=2,m=1,n=-1)$ [$\mu$Hz] & 84.353 & \cite{briquet_asteroseismic_2005}; id. by LPV  \\ 
 $\nu_7(\ell=2,m=2,n=-1)$ [$\mu$Hz] & 85.298 & \cite{briquet_asteroseismic_2005}; id. by LPV  \\ 
 T$_{\textrm{eff}}$ [K] & 22900 $\pm$ 900 & \cite{handler05} \\
 & 24000 $\pm$ 1000 & \cite{briquet_asteroseismic_2005} \\
 & 25000 $\pm$ 1000 & \cite{briquet09}; tertiary component included 
in the analysis \\
 $\log g$ [cm/s$^2$] & 4.00 $\pm$ 0.10 ; 4.10 $\pm$ 0.15 & \cite{briquet_asteroseismic_2005,briquet09} \\
 $\textrm{[Fe/H]}$ ; $Z$ & -0.04 $\pm$ 0.05 ; 0.0114 $\pm$ 0.0028 & \cite{briquet09} \\
 $v \sin i$ [km/s] & 30 -- 38.5 & \cite{briquet09}  \\
 
  \hline
 \hline
 \end{tabular}

 \end{table*}
     
    \label{section1}
    
    In this same table, we report two values for the metallicity. The first one is the iron abundance [Fe/H] by \cite{briquet09} and the second one is a mass fraction $Z$ that was derived by these authors. It is derived from the detailed abundances they could measure from the spectra of the star; then, they scaled down, based on [Fe/H], the abundances of undetermined elements, assuming they are distributed following the solar chemical mixture of \cite{grevesse98}. However, as shown in \cite{morel06}, \cite{norbert} or \cite{niem09}, the chemical mixture of solar-neighbourhood B-type stars is very similar to the solar chemical mixture as revised in AGS05, \citep{asplund}. Depending on the distribution adopted for the unknown elements, the metal mass fraction obtained can thus be impacted. We hence privileged the use of the [Fe/H] value as our indicator of the metallicity for $\theta$ Oph, since we adopted the AGS05 mixture in our models. \\

    \subsection{Theoretical models}
    The models used for asteroseismic modelling are computed with the Li\`{e}ge stellar evolution code CLES \citep{scuflaire_cles_2008}. Since the two stars are relatively slow rotators, the stellar models do not include rotational effects. The treatment of convection follows the mixing-length theory (MLT) prescription by \cite{cox68}, with its solar parameter calibrated at $\alpha_{MLT}$=1.8. The mixing in the overshoot region of the models is treated as instantaneous. The surface boundary conditions are obtained from Eddington's law ($T[\tau]$) for a grey atmosphere. The nuclear reaction rates are those of \cite{adelberger11} and the solar chemical mixture of AGS05 is adopted (see Sect.~\ref{section1}). Opacities corresponding to this chemical mixture are computed with the OPAL tables \citep{iglesias}, and the FreeEOS equation of state is used \citep{freeeos}.\\ 
    
    Oscillations are computed with the LOSC adiabatic oscillation code in its standard setting \citep[see details in][]{losc}. The optimisation method used for the asteroseismic fitting follows a local approach relying on the Levenberg-Marquardt algorithm, as implemented for use with CLES and LOSC \citep{salmon21}. 
    For $\beta$~CMa, the free parameters of the Levenberg-Marquardt optimisation are the stellar mass, the age and the initial chemical composition $Z/X_0$, with the core overshooting parameter fixed at $\alpha_{\textrm{ov}}=0.3$, the value giving the best results among the preliminary tests described in Sect.~\ref{sect:seismic_model}. For $\theta$~Oph, the additional seismic constraint allows a fourth free parameter, the core overshooting $\alpha_{\textrm{ov}}$.\\
        
    \subsection{\texorpdfstring{$\beta$}{b}~CMa asteroseismic solution}
    
    We started our seismic modelling using as constraints the three modes detected in \cite{mazumdar_asteroseismic_2006} and their identification, while trying to reproduce the subsolar metallicity of the star \citep{morel06}. We found that an overshooting of $\sim$~0.3 was necessary to reproduce the metallicity, but no stellar model was able to reproduce all three modes with the same accuracy. The identification of $\nu_3$ should be treated with caution, and we were not able to confirm it. From this first step, we achieved our modelling using the two modes with robust identification \citep[see][]{mazumdar_asteroseismic_2006} plus the effective temperature as a third constraint. We then obtained a solution with a good fit and reproducing the metallicity of the star. This asteroseismic solution (stellar parameters of the best-fit model) is given in Table~\ref{results-bcma}.\\
    
     \begin{table}
     \caption{Stellar parameters inferred from the asteroseismic modelling of $\beta$~CMa.}
     \label{results-bcma}
     \begin{center}
     \begin{tabular}{lc}
     \hline
    Stellar parameter & $\beta$~CMa model  \\
     \hline \hline
     Mass [M$_{\odot}$] & 13.2 $\pm$ 1.5  \\
     $Z/X_0$ & 0.017 $\pm$ 0.006  \\ 
     Age [Myr] & 14.6 $\pm$ 2.6 \\
     $\alpha_{\textrm{ov}}$ & 0.3 (fixed) \\
     R [R$_{\odot}$] & 10.5 $\pm$ 0.5 \\
     $\log g$ [dex] & 3.52 $\pm$ 0.02\\
      \hline
     \hline
     \end{tabular}
     \end{center}
     \end{table}

\begin{figure}
      \resizebox{\hsize}{!}{\includegraphics{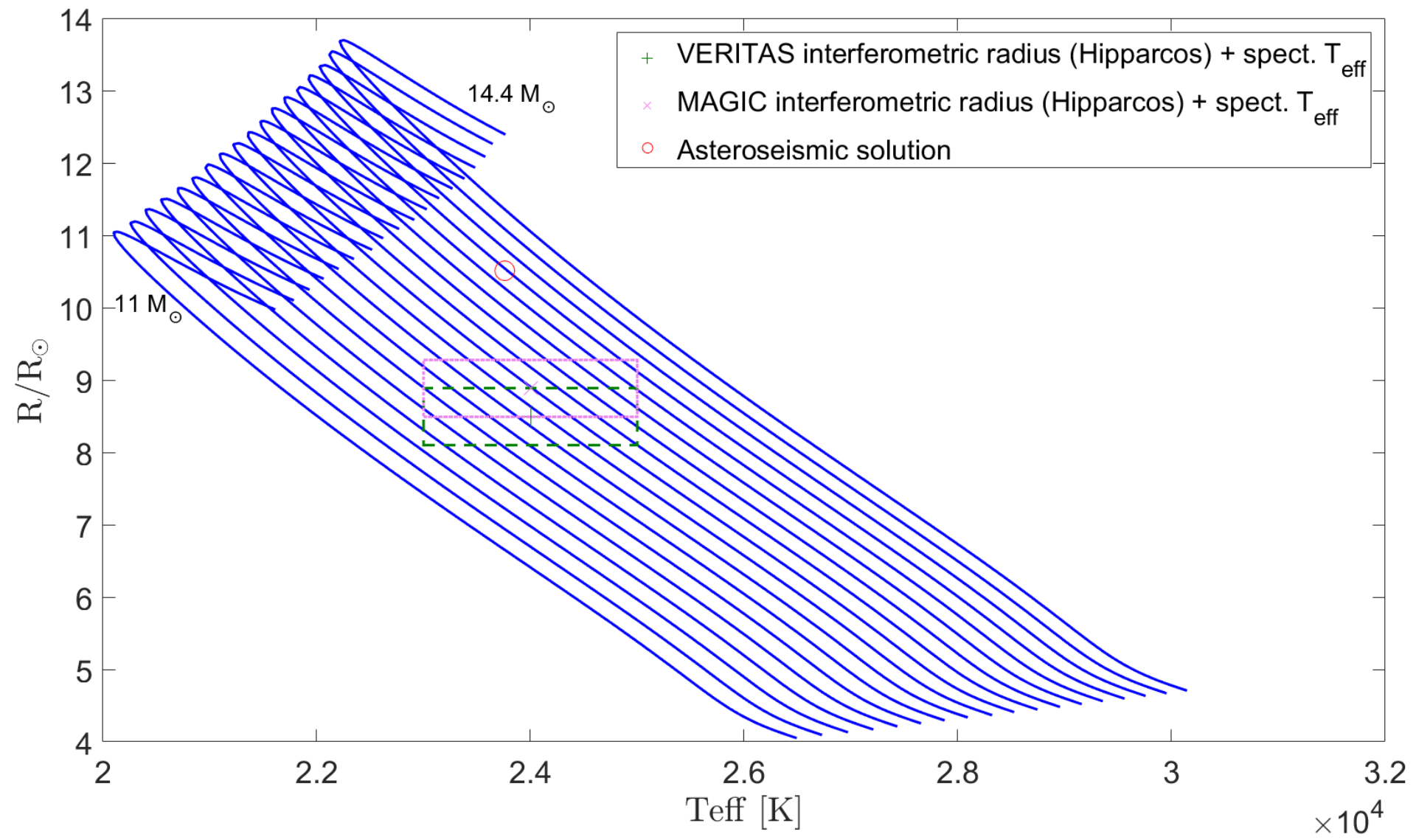}}
      \caption{Asteroseismic solution for $\beta$~CMa in a R- T$_{\textrm{eff}}$ diagram. The radius obtained from interferometry either from VERITAS or MAGIC-SII and T$_{\textrm{eff}}$ from \citet{morel06} are given with their error bars. Indicative main-sequence tracks for solar metallicity models between 11 and 14.4~M$_{\odot}$ (step of 0.2~M$_{\odot}$) are given in blue. The uncertainty on the asteroseismic radius, derived from a sensitivity analysis, is given in Sect.~5.4.}
      \label{fig:asteroseismic_bcma}
    \end{figure}

    In comparison to \cite{mazumdar_asteroseismic_2006}, which found a mass of 13.5$\pm$0.5~M$_{\odot}$, an age of $12.4\pm0.7$ Myr and $\sim Z/X_0=0.03$, our solution is in good agreement for the mass. However, their solution yielded a metallicity in disagreement with the spectroscopic one derived by \cite{morel06}. Our solution resolves this issue. It also reproduces the spectroscopic T$_{\textrm{eff}}$ from the same authors, as shown in figure~\ref{fig:asteroseismic_bcma}.

    Finally, we infer from the best-fit model a mean stellar density $\bar{\rho}=0.0161$~g\,cm$^{-3}$ and a radius $R=10.5$~R$_{\odot}$. Our optimisation does not provide the covariance matrix of the fitted parameters; we therefore estimate the uncertainty on the seismic radius from a sensitivity analysis. Since the observed frequencies essentially fix the mean density, with little dependence on the model input physics for low-order modes \citep{salmon22}, the radius follows from $R=(3M/4\pi\bar{\rho})^{1/3}$ and its relative uncertainty is dominated by that of the mass at fixed $\bar{\rho}$, $\delta R/R \simeq \frac{1}{3}\,\delta M/M \simeq 3.8\%$. Adding in quadrature the $\sim$2\% accuracy of the method established in hare-and-hound exercises when classical constraints are available \citep{salmon22}, we adopt $R=10.5\pm0.5$~R$_{\odot}$. The corresponding surface gravity is $\log g = 3.52\pm0.02$. In Fig.~\ref{fig:asteroseismic_bcma}, the predicted asteroseismic radius is larger than the two interferometric results; the significance and possible origins of this difference are examined in Sect.~\ref{sect:discussion}.\\

    \subsection{\texorpdfstring{$\theta$}{0}~Oph asteroseismic solution}\label{sect:theta_oph_seismic_solutions} 
    The asteroseismic modelling of $\theta$~Oph was done adopting one set of constraints, composed of the three frequencies with their identification and the metallicity [Fe/H]. The properties of the best-fit model are given in Table~\ref{results-thetaoph}.
    
     \begin{table}
     \caption{Stellar parameters inferred from the asteroseismic modelling of $\theta$~Oph.}
     \label{results-thetaoph}
     \begin{center}
     \begin{tabular}{lc}
     \hline
    Stellar parameter & $\theta$~Oph model\\
     \hline \hline
     Mass [M$_{\odot}$] & 8.6 $\pm$ 0.2 \\
     $Z/X_0$ & 0.022  \\ 
     Age [Myr] & 20.6 $\pm $3.2\\
     $\alpha_{\textrm{ov}}$ & 0.2 $\pm$ 0.1 \\
     R [R$_{\odot}$] &5.07 $\pm$ 0.11 \\
      \hline
     \hline
     \end{tabular}
     \end{center}
     \end{table}
     
    Our inferred stellar parameters are in line with previous stellar modelling of this target, except that we find a lower 
    value for the overshooting than the usual high range. Accordingly, our mass is slightly larger.

    \begin{figure}
      \resizebox{\hsize}{!}{\includegraphics{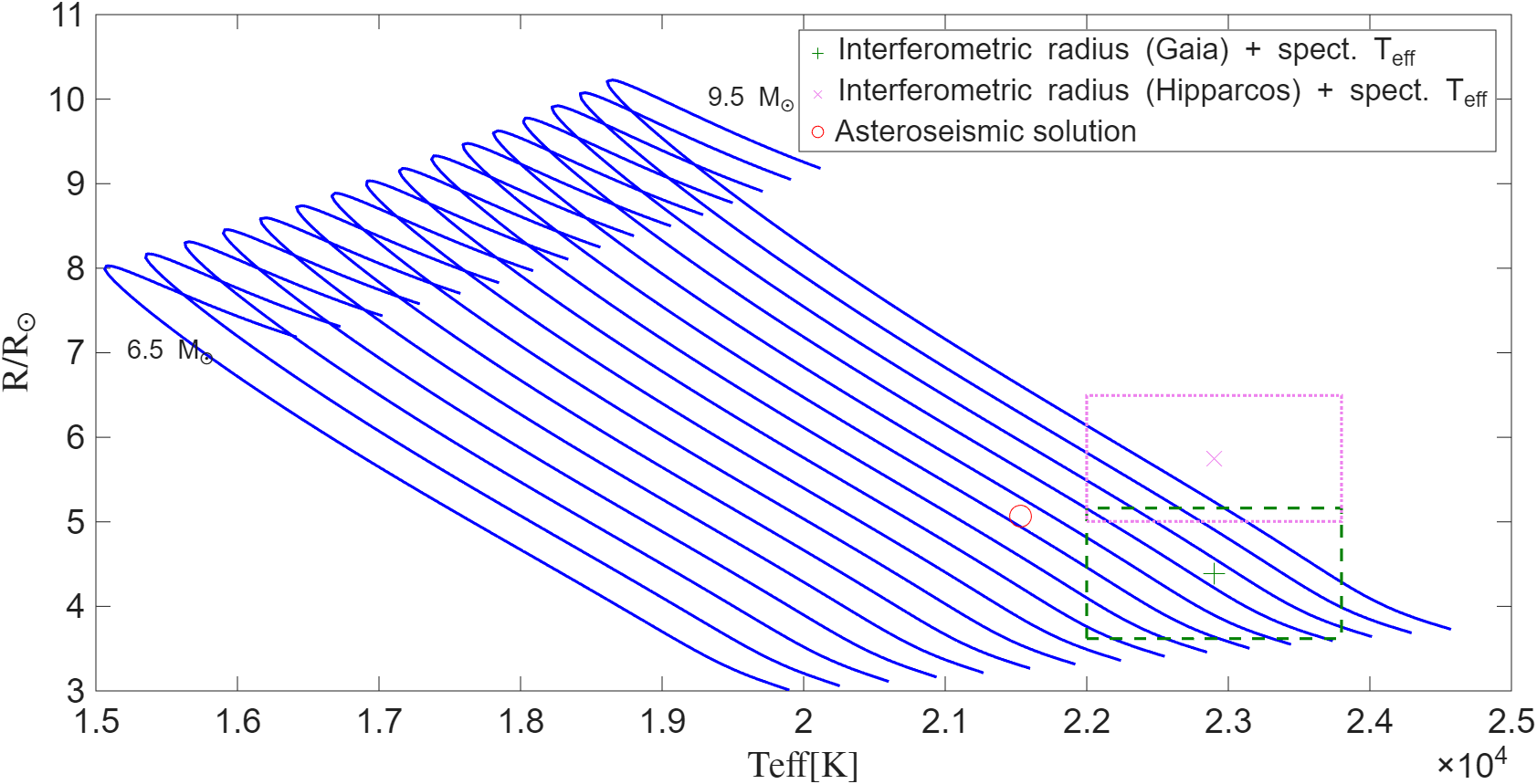}}
      \caption{Asteroseismic solution for $\theta$~Oph in a R- T$_{\textrm{eff}}$ diagram. The radius obtained from interferometry in this work and T$_{\textrm{eff}}$ from \citet{handler05} are given with their error bars. Indicative main-sequence tracks for solar metallicity models between 6.5 and 9.5~M$_{\odot}$ (step of 0.2~M$_{\odot}$) are given in blue. The uncertainty on the asteroseismic radius, derived from a sensitivity analysis, is given in Sect. \ref{sect:theta_oph_seismic_solutions}.}
      \label{fig:asteroseismic_thetaoph}
    \end{figure}

    The effective temperature of our solution, T$_{\textrm{eff}}=$21534~K, does not reproduce the spectroscopic values of \cite{briquet_asteroseismic_2005}. It is marginally compatible with the narrow-band photometric one \citep{handler05}, as illustrated in figure~\ref{fig:asteroseismic_thetaoph}. This disagreement with the spectroscopic temperature was also observed in the previous asteroseismic studies of this star. The discrepancy is likely due to the complexity of the system, as the companion of the speckle binary contaminates the spectral lines by 20\%. We finally estimate from the best-fit model $\bar{\rho}=0.092$~g\,cm$^{-3}$ and $R=5.07\pm0.11$~R$_{\odot}$, where the uncertainty is derived with the same sensitivity approach as for $\beta$~CMa (dominated here by the $\sim$2\% method accuracy, given the small mass uncertainty). This radius is in agreement with the value derived based on the interferometric value we derived (Table~\ref{table:results}), though only at the end of the error intervals. It must be noted that the parallax errors dominate the radius derived from interferometry for the Hipparcos result. Additionally, the Hipparcos ($7.48\pm0.17$~mas) and Gaia DR2 ($9.79\pm1.18$~mas) parallax values are in disagreement at the $1.9\sigma$ level. We discuss this aspect in more detail hereafter

\section{Discussion}\label{sect:discussion}

    Asteroseismology enables the determination of stellar radii for $\beta$~Cephei stars with remarkable accuracy. \citet{salmon22} have investigated in thorough detail how the asteroseismic determination of this parameter depends on either the number of detected (axisymmetric) frequencies or the physics of the stellar models. Independently of this physics, with the sole constraint of a set of at least four identified frequencies, the accuracy on the radius is expected to be under 1\%. For $\beta$~CMa and $\theta$~Oph, the seismic constraints were fewer in number, but we could make use of classical constraints (T$_{\textrm{eff}}$, metallicity). In this case, the same authors demonstrated that the quality of the asteroseismic determination remained very accurate, and one could expect a precision (error bar) of 1.5-2\%, again independently of the physics of the models. This is due to the physical nature of the pulsations of $\beta$~Cephei stars, whose modes are very sensitive to the mean stellar density. Thus, as it depends on the inverse cube of the radius, this latter parameter is strongly constrained by asteroseismology.

    \begin{figure}
      \resizebox{\hsize}{!}{\includegraphics{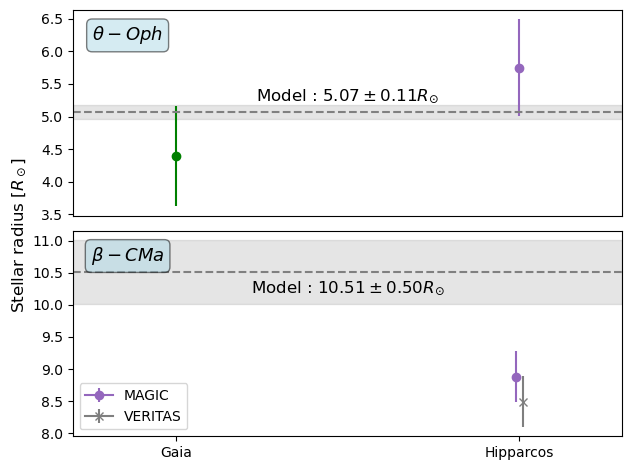}}
      \caption{Full comparison of the different results obtained in this work. 
      The uncertainty on the radius for $\theta$ Oph is larger than that of $\beta$ CMa because no ZBC calibration was used. 
      The MAGIC-SII result is compatible with the VERITAS one.}
      \label{fig:MAGIC_result_comp}
    \end{figure}

    Regarding the interferometry analysis, the $\theta$ Oph radius is $4.4\pm0.8$~R$_{\odot}$ with Gaia parallax and $5.7\pm0.7$~R$_{\odot}$ with the Hipparcos one. In both cases, the relative error on the radius exceeds $12.7\%$  because no ZBC could be used for this star, as explained earlier. For the same reason, the asteroseismic model is consistent with both parallax assumptions.

    Concerning $\beta$~CMa, the radius derived from the MAGIC-SII angular diameter and the Hipparcos parallax, $8.9\pm0.4$~R$_{\odot}$, differs from the asteroseismic prediction of $10.5\pm0.5$~R$_{\odot}$ at the $2.5\sigma$ level once the uncertainty of the seismic radius is accounted for. The absence of a Gaia parallax measurement makes it difficult to determine whether the difference originates from Hipparcos or from the modelling. However, \cite{abeysekara_demonstration_2020} measured the angular diameter of $\beta$ CMa using four VERITAS telescopes, obtaining a value of $0.523\pm0.017$~mas, compatible with the MAGIC-SII result and corresponding to $8.5\pm0.4$~R$_{\odot}$, i.e.\ a $3.1\sigma$ difference with the seismic prediction. In both cases, a small error is expected as the radius was derived assuming a uniform disk, while a limb-darkened model is more representative; however, as seen in \cite{magic_2024} and \cite{abeysekara_demonstration_2020}, the corresponding radius increase is only of a few percent and cannot explain the difference.
    
    The origin of this tension deserves a balanced examination, since neither the parallax nor the spectroscopic constraints nor the seismic modelling can be excluded a priori. A first, quasi-model-independent test is obtained by combining the seismic mean density with the interferometric radius: since low-order $\beta$~Cep pulsation frequencies constrain $\bar{\rho}$ with very little sensitivity to the model input physics \citep{salmon22}, taking the Hipparcos-based radii at face value implies $M=\bar{\rho}\,\frac{4\pi}{3}R^{3} = 8.1\pm1.1$~M$_{\odot}$ (MAGIC-SII) or $7.0\pm1.0$~M$_{\odot}$ (VERITAS). Such masses are incompatible with all spectroscopic and evolutionary estimates for this B1\, II-III star, which lie between 12 and 14~M$_{\odot}$ \citep{mazumdar_asteroseismic_2006,logg15,logg25}. Percent-level systematics of the stellar models, as quantified in the input-physics tests of \cite{salmon22}, are therefore quantitatively insufficient to resolve the tension: within the adopted mode identification, the seismic mean density would have to be in error by $\sim$60\%, i.e.\ the measured frequencies by $\sim$25\%, which is excluded. The same argument applies to the rotational correction of the non-axisymmetric mode: the rotational kernel of a low-degree $g_1$ mode in a $\beta$~Cep star has most of its amplitude in the envelope 
    \citep{salmon22b}, except in evolved stars where it can peak in the $\mu$-gradient zone, as in HD~192575 \citep{burssens_calibration_2023} and possibly in $\beta$~CMa, so that the surface rotation is an approximate proxy of the rotation it senses; empirically, in HD~129929 the $g_1(\ell=2)$ and $p_1(\ell=1)$ splittings agree to 0.2\% despite a core-to-surface rotation contrast of $\sim$3.9 \citep{salmon22b}, and the majority of $\beta$~Cep stars show envelope-to-surface rotation ratios between 0.3 and 2 \citep{fritzewski25}. Even a 100\% error on the $\sim$1~$\mu$Hz correction applied here would shift the inferred radius by $\lesssim$1.5\%, far below the observed $\sim$15\% difference.
    
    A second, so far overlooked element comes from the spectroscopic literature itself: the published determinations of the surface gravity of $\beta$~CMa split into two mutually incompatible groups (Table~\ref{table-logg-bcma}). On the one hand, \cite{logg96}, \cite{morel06} and \cite{logg23} report $\log g$ in the range 3.4--3.54, the most recent and most precise homogeneous determination being $3.54\pm0.04$ \citep{logg23}; on the other hand, \cite{logg15} obtained $3.78\pm0.08$, the two groups differing at the $2.7\sigma$ level. This dichotomy maps exactly onto the two radius solutions. Our asteroseismic solution, which used only $T_{\textrm{eff}}$ (and not $\log g$) as a classical constraint, implies $\log g = 3.52\pm0.02$, in remarkable a posteriori agreement with the first group. Conversely, the interferometric radius combined with the Hipparcos parallax implies $\log g \simeq 3.62$--$3.70$ for masses between 12 and 14~M$_{\odot}$, consistent with the second group and with the evolutionary estimates of \cite{logg15} ($M=12.0^{+0.3}_{-0.7}$~M$_{\odot}$, $R=7.4^{+0.8}_{-0.9}$~R$_{\odot}$ and $M=12.6^{+0.4}_{-0.5}$~M$_{\odot}$, $R=8.2\pm0.6$~R$_{\odot}$). We note that the recent determination $\log g = 3.70\pm0.08$ by \cite{logg25} is itself derived from the VERITAS angular diameter and the Hipparcos parallax, and is therefore consistent with the second group by construction. The tension between the seismic and parallax-based radii thus mirrors a pre-existing, unresolved conflict between independent spectroscopic analyses of this star, and cannot be resolved internally with the present data.
    
    The weight of the parallax term in this comparison deserves scrutiny. The adopted value, $6.62\pm0.22$~mas, is that of the van Leeuwen (2007) re-reduction; the original Hipparcos catalogue gives $6.53\pm0.66$~mas \citep{perryman97}, differing by only 0.09~mas, but with an uncertainty three times larger. Adopting the original catalogue value, the interferometric radius becomes $9.0\pm0.9$~R$_{\odot}$ (MAGIC-SII) and $8.6\pm0.9$~R$_{\odot}$ (VERITAS), and the tension with the seismic prediction drops to $1.4\sigma$ and $1.8\sigma$, respectively, while the parallax required by the seismic solution lies within $1.3\sigma$ of the original catalogue value. The significance of the discrepancy thus rests on the formal precision claimed by the re-reduction for a $V\simeq2$ star, in the brightness regime where the Hipparcos re-reduction is most delicate.

    The tension thus points to an underestimated uncertainty, or a bias, of the re-reduced Hipparcos parallax, possibly related to the extreme brightness of $\beta$~CMa, a scenario supported both by the comparison of the two Hipparcos reductions above and by the a posteriori agreement of the seismic solution with three independent spectroscopic $\log g$ determinations. By reverse calculation, matching the seismic radius requires a parallax of $5.6\pm0.3$~mas (MAGIC-SII) or $5.4\pm0.3$~mas (VERITAS), i.e.\ $2.7\sigma$ away from the re-reduced value but only $1.3\sigma$ and $1.6\sigma$ from the original catalogue value.
    This difference seems to be compatible with the average systematic offsets between Hipparcos and interferometry ($\sim1.3$~mas) mentioned in \cite{pan_systematic_2007}.
    We note that the alternative --- that the high-$\log g$ spectroscopic solution is the correct one, together with the Hipparcos-based radius --- cannot be formally excluded by the parallax alone, although our asteroseismic log g disfavours the high-log g value of \cite{logg15} (see Table \ref{table-logg-bcma}).
    
\begin{table*}[h]
    \caption{Published determinations of the surface gravity of $\beta$~CMa, compared with the values implied by the two radius solutions.}
    \label{table-logg-bcma}
    \centering
    \begin{tabular}{l l l}
    \hline
    $\log g$ [dex] & Reference & Method \\
    \hline\hline
    3.4--3.5      & \cite{logg96} & EUVE spectrum + model atm. \\
    3.50 $\pm$ 0.15 & \cite{morel06} & NLTE spectroscopy \\
    3.54 $\pm$ 0.04 & \cite{logg23} & FASTWIND, homogeneous survey \\
    3.78 $\pm$ 0.08 & \cite{logg15} & spectroscopy \\
    3.70 $\pm$ 0.08 & \cite{logg25} & $\theta_{\rm UD}$(VERITAS) + $\pi$(Hip)$^{(a)}$ \\
    \hline
    3.52 $\pm$ 0.02 & this work & asteroseismology \\
    3.62--3.70      & this work & $\theta_{\rm UD}$ + $\pi$(Hip), $M$=12--14~M$_{\odot}$ \\
    \hline
    \hline
    \end{tabular}
    \tablefoot{$^{(a)}$ Derived from the interferometric angular diameter and the Hipparcos parallax; not independent of the parallax-based radius.}
\end{table*}

    The angular diameter uncertainties are expected to decrease in the next few months or years with the commissioning of the four Large-Sized Telescopes (LSTs). Their larger collecting area, higher quantum efficiency and combination of baselines will help to increase the amount of data, reducing the overall uncertainty. Additionally, the Gaia parallax is expected to be updated in the next data release (DR4), potentially improving its accuracy on bright stars \footnote{\url{https://www.cosmos.esa.int/web/gaia/data-release-4}}.
    
    Parallax uncertainty reduction is critical because the stellar radius uncertainty is proportional to $\delta\bar{\omega}/\bar{\omega}$. 
    The two stars studied in this work are bright and are at the edge of Gaia's bright limit (magnitude $G\approx3$ for the five-parameter solution \footnote{See \url{https://www.cosmos.esa.int/web/gaia/dr2} for more details.}). As computed in Section \ref{sect:results}, with the current amount of data, the uncertainty is dominated by the MAGIC-SII uncertainty, but more SII measurements are being taken. 
    
    LSTs, which can also take interferometry data, can significantly increase the amount of data and allow observations on longer baselines, which allow to resolve smaller angular scales. This could, with reasonable observation time, reduce the uncertainty on the angular diameter by a factor of $\sim10$, reaching $\sim1\%$ uncertainty. With such error bars and low systematic uncertainties on the parallax (\textit{ie.}, similar to the current statistical parallax uncertainty), the uncertainty on the $\theta$ Oph radius would be $12.1\%$ assuming Gaia parallax, while the uncertainty on $\beta$ CMa would drop to $3.4\%$.
    Similarly, assuming an uncertainty of $1\%$ on both the angular diameter and the model, one can infer a $1.4\%$ uncertainty in the parallax. 
    
    If confirmed by other SII measurements with additional data and asteroseismic simulations, the stellar diameter could serve as a calibration method for parallax measurements of bright $(G\approx3)$ stars. Still, it could also be used to constrain asteroseismic models.

\section{Conclusion}\label{sect:conclusion}
    
    Overall, the analyses we performed, especially the one on $\theta$ Oph,  are very promising. The asteroseismic modelling predicted a stellar radius of $5.07\pm0.11$~R$_\odot$ while the MAGIC-SII result using Gaia parallax is $4.4\pm0.8$~R$_\odot$, and $5.7\pm0.7$~R$_\odot$ using Hipparcos parallax. 
    
    For $\theta$ Oph, the lack of a zero baseline correlation value increases the uncertainty on the stellar diameters, which explains why the two parallaxes used for this star (although being incompatible) lead to results that are compatible with the model.  
    
    Regarding $\beta$ CMa, the asteroseismic modelling predicts a radius of $10.5\pm0.5$~R$_\odot$, in tension at the $2.5\sigma$ level with the MAGIC-SII result of $8.9\pm0.4$~R$_\odot$ obtained with the Hipparcos parallax; a similar study from the VERITAS collaboration is in good agreement with the MAGIC-SII angular diameter and yields a $3.1\sigma$ tension with the asteroseismic prediction. We showed that percent-level systematics of the seismic modelling, including the treatment of the rotational correction of non-axisymmetric modes, cannot account for this difference, and that it mirrors a pre-existing conflict between published spectroscopic determinations of the surface gravity of the star. The tension thus points to a bias of the Hipparcos parallax, whose value should then be $5.6\pm0.3$~mas, a scenario supported by the a posteriori agreement of the seismic solution with three independent spectroscopic $\log g$ determinations. Gaia DR4 and forthcoming SII measurements with the CTAO-North LSTs will settle the issue.
    The results are promising and justify additional data taking to reach $\sim1\%$ relative uncertainty on the angular diameter, which could lead to smaller uncertainty on the radii down to $\sim3\%$, and uncertainty on the parallax to $1.4\%$.

\begin{acknowledgements}
We would like to thank the Instituto de Astrof\'{\i}sica de Canarias for the excellent working conditions at the Observatorio del Roque de los Muchachos in La Palma. The financial support of the German BMFTR, MPG and HGF; the Italian INFN and INAF; the Swiss National Fund SNF; the grants PID2022-136828NB-C41, PID2022-137810NB-C22, PID2022-138172NB-C41, PID2022-138172NB-C42, PID2022-138172NB-C43, PID2022-139117NB-C41, PID2022-139117NB-C42, PID2022-139117NB-C43, PID2022-139117NB-C44, CNS2023-144504 funded by the Spanish MCIN/AEI/ 10.13039/501100011033 and "ERDF A way of making Europe"; the Indian Department of Atomic Energy; the Japanese ICRR, the University of Tokyo, JSPS, and MEXT; the Bulgarian Ministry of Education and Science, National RI Roadmap Project D01-135/16.07.2025 and the Academy of Finland grant nr. 320045 is gratefully acknowledged. This work has also been supported by Centros de Excelencia ``Severo Ochoa'' y Unidades ``Mar\'{\i}a de Maeztu'' program of the Spanish MCIN/AEI/ 10.13039/501100011033 (CEX2019-000918-M, CEX2021-001131-S, CEX2024-001441-S, CEX2025-001609-S), by AST22\_00001\_9 with funding from NextGenerationEU funds and by the CERCA institution and grants 2021SGR00426, 2021SGR00607 and 2021SGR00773 of the Generalitat de Catalunya; by the Croatian Science Foundation (HrZZ) Project IP-2022-10-4595 and by the University of Rijeka Project uniri-mzi-25-3 funded by the European Union - NextGenerationEU; by the Deutsche Forschungsgemeinschaft (SFB1491) and by the Lamarr-Institute for Machine Learning and Artificial Intelligence; by the Polish Ministry of Science and Higher Education grant No. 2025/WK/04; by the European Union (ERC/MicroStars, 101076533 and INFRA/ACME, 101131928); and by the Brazilian MCTIC, the CNPq Productivity Grant 309053/2022-6 and FAPERJ Grants E-26/200.532/2023 and E-26/211.342/2021.
\\

\textbf{Author Contribution}
The MAGIC-SII analysis part was led by GK and RW. Comments, cross-checks, and reviews from TH, ACS, IJM, JC and D.Cuenda-Muñoz were highly appreciated. Sébastien Salmon from the stellar group of the Geneva Observatory provided the asteroseismic modelling part.

The rest of the authors have contributed in one or several of the following ways: design, construction, maintenance, and operation of the instrument(s) used to acquire the data; preparation and/or evaluation of the observation proposals; data acquisition, processing, calibration and/or reduction; production of analysis tools and/or related Monte Carlo simulations; overall discussions about the content of the draft, as well as related refinements in the descriptions.

\end{acknowledgements}

\bibliographystyle{aa}
\bibliography{bibliography}
\end{document}